# The effect of simultaneous substitution on the electronic band structure and thermoelectric properties of Se-doped $Co_3SnInS_2$ with the Kagome lattice

Masaya Fujioka[1], Taizo Shibuya[2], Junya Nakai[3], Keigo Yoshiyasu[3], Yuki Sakai[3], Yoshihiko Takano[1], Yoichi Kamihara[3] and Masanori Matoba[3]

[1]National Institute for Materials Science, 1-2-1 Sengen, Tsukuba, Ibaraki 305-0047, Japan

[2]Department of Mechanical Engineering, Keio University, 3-14-1 Hiyoshi, Yokohama 223-8522, Japan

[3]Department of Applied Physics and Physico-Informatics, Keio University, 3-14-1 Hiyoshi, Yokohama 223-8522, Japan

Corresponding author: Masaya Fujioka

E-mail : FUJIOKA.Masaya@nims.go.jp

Abstract

The thermoelectric properties and electronic band structures for Se-doped $Co_3SnInS_2$ were examined. The parent compound of this material ($Co_3Sn_2S_2$) has two kinds of Sn sites (Sn1 and Sn2 sites). The density functional theory (DFT) calculations show that the indium substitution at the Sn2 site induces a metallic band structure, on the other hand, a semiconducting band structure is obtained from substitution at the Sn1 site. However, according to the previous reports, since the indium atom prefers to replace the tin atom at the Sn1 site rather than the Sn2 site, the resistivity of $Co_3SnInS_2$ shows semiconducting-like behavior. In this study we have demonstrated that metallic behavior and a decrease in resistivity for Se-doped $Co_3SnInS_2$ occurs without suppression of the Seebeck coefficient. From the DFT calculations, when the selenium content is above 0.5, the total crystallographic energy shows that a higher indium occupancy at Sn2 site is more stable. Therefore, it is suggested that the selenium doping suppress the site preference for indium substitution. This is one of the possible explanations for the metallic conductivity observed in Se-doped $Co_3SnInS_2$

1. INTRODUCTION

Compounds with the general formula $M_3A_2Ch_2$ (M = Co, Ni, Rh and Pd; A = Sn, Pb, In, Tl, and Bi; Ch = S and Se) [1-9] can be classified accordingly by the component A. When A ≠ Bi, $M_3A_2S_2$ shows the shandite-type structure such as the mineral $Ni_3Pb_2S_2$. When A = Bi, $M_3Bi_2S_2$ shows the parkerite-type structure such as the mineral $Ni_3Bi_2S_2$. Figure 1 (a) shows the structure of the shandite-type $Co_3Sn_2S_2$ [4]. This material has a Kagome lattice composed of cobalt as shown in figure 1 (b). Some reports show a magnetic transition from paramagnetic to ferromagnetic at 177 K [10, 11, 12]. It is also reported that the easy axis of magnetization is perpendicular to the Kagome lattice [12]. Furthermore, $Co_3Sn_2S_2$ shows metallic resistivity (around 1 mΩcm at 300 K) and a large negative Seebeck coefficient (around - 50 μV/K at 300 K) [7, 10, 11]. The electronic structure of $Co_3Sn_2S_2$ indicates half metallic ferromagnetism [12, 13].

Indium substituted $Co_3Sn_2S_2$ ($Co_3Sn_{2-x}In_xS_2$) has been already studied [7, 14-19] and a large change in the physical properties depending on the amount of indium content has been confirmed. With increasing indium content from x = 0 to x = 1.0, resistivity increases and the Seebeck coefficient enhances to a large negative value. In contrast, with further increasing indium content from x = 1.0 to 2.0, the resistivity decreases and Seebeck coefficient is suppressed to a small negative value. In addition, the ferromagnetism observed in $Co_3Sn_2S_2$ completely disappears, when x is above 1.0. Therefore, x = 1.0 is a significant point where one is able to characterize its physical properties, in regards to the indium content, namely resistivity and Seebeck coefficient. The power factor should be enhanced when the indium content is changed slightly from x = 1.0. Actually, when

x = 0.85, the maximum figure of merit ZT (= 0.2) at 300 K was reported [18]. The $ZT$ for a thermoelectric material is determined by the equation ($ZT = \alpha^2 \cdot T/(\kappa \cdot \rho)$), where $\rho$ is the electrical resistivity, $\alpha$ is the Seebeck coefficient, $\kappa$ is the thermal conductivity and $T$ is temperature. To achieve a large figure of merit, these physical parameters must all be improved. In this case, optimization of the amount of indium is thought to induce low resistivity and a high Seebeck coefficient.

Meanwhile, the change in lattice parameters with x does not follow Vegard's law. This behavior suggests site preference for $Co_3Sn_{2-x}In_xS_2$ [15-19] The structure of $Co_3Sn_2S_2$ contains two kinds of Sn sites. One site, which is called the Sn1 site, is located between layers of cobalt, and the other site, which is called the Sn2 site, is located on the same layer as the cobalt as shown in figure 1 (a), (c). In a previous study, electronic band structures of two kinds of $Co_3SnInS_2$ were calculated [16]. These $Co_3SnInS_2$ were distinguished by which Sn sites are occupied by indium. When all of the Sn1 sites are occupied, the electronic state calculations show a semiconducting band structure with a narrow band gap. Meanwhile, when all of the Sn2 sites are occupied, it shows a metallic band structure. Considering the physical properties of $Co_3SnInS_2$ with semiconductor-like behavior, it is known that the tin on the Sn1 site is easier to be substituted than that of the Sn2 site. Furthermore, in recent research, the sight preference of this material is discussed by $^{119}$Sn Mössbauer spectroscopy and DFT calculations [19]. When x = 1.0, the tin occupancy of Sn1 : Sn2 is given as 3 : 7. This result is consistent with the above mentioned results.

In this research, The Indium amount is also fixed at x = 1.0, because at this value, a significant change in the physical properties of $Co_3SnInS_2$ occurs. Additionally, simultaneous substitution of selenium is performed from y = 0.0 to 1.0. The resistivity, Seebeck coefficient, thermal conductivity and DFT calculations were performed to discuss the effect of selenium substitution.

2. EXPERIMENTAL SECTION

2.1 Sample preparation

Polycrystalline $Co_3SnInS_{2-y}Se_y$ was synthesized by a solid-state reaction method. Powders of Sn (Kojundo Chemical; 99.99%), In (Aldrich; 99.999%), Co (Aldrich; 99.995%), S (Kojundo Chemical; 99.99%) and Se (Kojundo Chemical; 99.99%) were ground together with a mortar and pestle prior to sealing in an evacuated quartz tube. The mixture was sintered at 900 °C for 1 day and cooled to room temperature. Then, the mixture was removed from the quartz tube and reground, compressed into pellets and sealed again into a second quartz tube. They were resintered at 650 °C for 5 days. $Co_3SnInS_{2-y}Se_y$ (y = 0.2, 0.4, 0.6, 0.8, 1.0) were prepared.

2.2 Measurements

These Polycrystalline samples were characterized by X-ray diffraction (XRD) (Bruker AXS D8 ADVANCE) using Cu Kα radiation. The positions of all obtained peaks are calibrated by standard Si peaks (RMS 640c, NIST). The lattice parameters were estimated from 2θ and Miller index (hkl) using the least squares method. The elemental compositions of the samples were investigated by energy dispersive X-ray spectrometry (EDX) (HITACHI SU-70). The temperature dependence of resistivity ($\rho$), Seebeck coefficient ($\alpha$), thermal conductivity ($\kappa$) were measured from 300 K to 30 K. All samples were formed into rectangles with dimensions of around $1 \times 1.5 \times 5$ mm$^3$. The sintered densities of all samples (y = 0.0, 0.2, 0.4, 0.6, 0.8 and 1.0) were estimated to be 63, 75, 75, 76, 77 and 67 % of the theoretical densities respectively. The measurement of resistivity was performed by a standard four-probe technique using Au electrodes. The Seebeck coefficient was measured by a four-probe technique using constantan and copper wires with a 25 μm diameter. Heat was applied so that a thermal gradient between the thermoelectric couples was stable at around 1 – 2 K. This type of thermoelectric couple was also used for the measurement of thermal conductivity. A steady state technique was applied for this measurement. Electrolytic Fe [20] was used as standard material for the calibration of the thermal conductivity. All measurements were performed by self-produced instruments with a cooling and compression machine (RDK-101D and CAN-11B, NAGASE). The thermoelectric power factor ($\alpha^2/\rho$) and figure of merit ($ZT$) were calculated from these results.

2.3 Calculation methods

For the DFT calculations, we used a plane wave based pseudopotential code, VASP [21,22]. The exchange correlation potential was approximated within the generalized gradient approximation by the Perdew-Becke-Ernzerhof (PBE) method [23]. The plane wave cut-off energy was 500 eV and the core-valence interaction was described by the Projector Augmented Wave approach (PAW). A Gaussian smearing with a width of sigma = 0.01 eV for the occupation of the electronic level was used. In order to investigate the electronic states affected by the existence of indium at the Sn2 sites, we focused on the composition of $Co_3SnInS_2$. The system was modeled by a 2×2×2 supercell. This amounts to 24 Sn, 24 In, 72 Co and 48 S atoms. We calculated the following three types: All (24 of 24) the indium atoms at Sn1 sites, 92% (22 of 24) of the indium atoms at Sn1 sites and the remaining 8% at the Sn2 sites, 75% (18 of 24) of the indium atoms at Sn1 sites and the remaining 25% at the Sn2 sites. They are labeled as ($Sn1_{In100}$), ($Sn1_{In92}$) and ($Sn1_{In75}$) respectively. Furthermore, for ($Sn1_{In92}$) and ($Sn1_{In75}$) we calculated y = 0.125 (3/48), 0.50 (12/48), 1.0 (24/48) in $Co_3SnInS_{2-y}Se_y$ to reveal the effect of selenium substitution. Lattice constants were experimentally determined, i.e. a = 0.5311 nm and c = 1.3476 nm for y = 0.0, a = 0.5320 nm c = 1.3498 nm for y = 0.125, a = 0.5339 nm and c = 1.3609 nm for y = 0.50 and a = 0.5352 nm c = 1.3635 nm for y = 1.0 were used. In the super cell, each atom of indium and selenium is placed at the furthest site from the position of the other indium and selenium atoms. All the atoms were relaxed until the Hellmann-Feynmann forces had become smaller than 0.05 eV/A. The Brillouin zone was sampled by a 3×3×1 Monkhorst-Pack grid. Furthermore, the phase stability of ($Sn1_{In92}$) and ($Sn1_{In75}$) with y = 0, 0.125, 0.50 and 1.0 were checked from the total energy estimated from the DFT calculations.

3. RESULTS and DISCUSSION

3.1 Sample characterization

For the XRD results of $Co_3SnInS_{2-y}Se_y$, with the exception of y = 1.0, almost all the diffraction peaks are assigned to the $Co_3SnInS_{2-y}Se_y$ phase. The XRD pattern of y = 1.0 contained a small amount of impurity phases (In$_6$Se$_7$) as shown in figure 2. Figure 3 and the inset show the lattice parameters of the *a*- and *c*-axes and the detected composition of selenium and indium as a function of y for Co$_3$SnInS$_{2-y}$Se$_y$ (0 < y < 1.0) respectively. The gradual increase in the actual selenium composition is confirmed from the change of the lattice parameters and the EDX results. On the other hand, actual In composition is slightly lower than that of the nominal composition. However, it is almost the same between y = 0 and 0.8. When y = 1.0, the formation of In$_6$Se$_7$ is thought to be the reason for the decrease in the indium composition. In addition, the samples are identified by their nominal compositions in this article.

3.2 Thermoelectric properties

Figure 4 (A) shows the temperature dependence of resistivity. The samples with selenium content from y = 0.0 to y = 0.4 shows negative d$\rho$/dT. This behavior is semiconducting-like. However, a drastic decrease of resistivity is observed for y = 0.6. With increasing y from 0.6 to 1.0, d$\rho$/dT shows a positive value. Figure 4 (B) shows the temperature dependence of the Seebeck coefficient. These results also show the different behaviors between y = 0.4 and y = 0.6. The maximum value of the Seebeck coefficient is - 171.72 µV/K at 240 K when y = 0.6. The peak negative Seebeck coefficient shifts to a higher temperature with greater selenium content than y = 0.6. Figure 4 (C) shows the temperature dependence of thermal conductivity. Although the results of resistivity and the Seebeck coefficient indicate a drastic change, thermal conductivity shows almost the same behavior in all samples. From the results of figure 4 (A) and (B), the power factor as a function of y is calculated as shown in figure 4 (D). With increasing y, the power factors clearly increase. The maximum value of the power factor is $6.40 \times 10^{-4}$ W/K$^2$m for y=1.0 at 300 K. Figure 4 (E) shows the ZT as a function of temperature for each sample. The tendency to increase can be observed with increasing selenium content. Maximum ZT reaches 0.067 at 300 K for y = 0.8. This improvement results from the drastic decrease in resistivity without suppression of the Seebeck coefficient induced by selenium substitution. Additionally, in comparison with ref. 18, there is a large difference in the thermal properties of Co$_3$SnInS$_2$. This is largely due to differences in sample density and composition.

3.3 Electronic band structure

Weihrich et al reported electronic band structure calculations assuming that indium occupies the either the Sn1 or Sn2 sites of Co$_3$SnInS$_2$. They reported a semiconducting band structure for the former and a metallic band structure for the latter as described above [16]. However, in actuality, indium is substituted for tin at both the Sn1 and Sn2 sites. To investigate the relevance between the electronic band structure and the physical properties of Co$_3$SnInS$_2$, a more detailed indium occupancy of each Sn site should be considered. Therefore, in this study, the three types of indium occupancy were adopted as (Sn1$_{In100}$), (Sn1$_{In92}$) and (Sn1$_{In75}$), as described above.

The densities of states (DOS) for these three types are shown in the figure 5 (A), (B) and (C). The band gap observed in figure 5 (A) decreases with an increasing amount of indium substitution for the

Sn2 site. Finally, in figure 5 (C), the band gap disappears. In practice, when x = 1.0, the behavior of resistivity has a negative $d\rho/dT$. We suggest that the actual indium occupancy of Sn1 site in $Co_3SnInS_2$ is thought to be more than 75 % due to the semiconducting behavior.

Figure 6 shows the DOS of $Co_3SnInS_{2-y}Se_y$. Since the resistivity changes from a semiconducting to a metallic behavior between y = 0.4 and 0.6, we expected an increase in DOS at the Fermi level with an increasing amount of selenium substitution. However, for $Sn1_{In92}$, a drastic change of DOS at the Fermi level could not be observed, as shown in figure 6 (A), (B) and (C). There is no evidence for orbital levels of selenium at the Fermi level from the electronic state calculations. This means that selenium substitution is not directly related to the increase of DOS at the Fermi level.

Table 1 shows the lattice parameters and the difference in total energy between $Sn1_{In92}$ and $Sn1_{In75}$ ($E(Sn1_{In92})-E(Sn1_{In75})$) for each amount of selenium substitution. From this table, it is suggested that a higher indium occupancy at Sn2 site is more stable when increasing selenium concentration. For instance, when y = 0 and 0.125, $E(Sn1_{In92})-E(Sn1_{In75})$ denotes the positive value. It means $Sn1_{In92}$ is more stable than $Sn1_{In75}$. On the other hand, when y = 0.5 and 1.0, the opposite result is observed. $Sn1_{In75}$ is more stable. Therefore, it is suggested that selenium substitution gives rise to an increase in the indium occupancy at the Sn2 site. In fact, when y = 0.5, $Sn1_{In75}$ shows a metallic electronic band structure as shown in figure 6 (D). This result agrees with the behavior of resistivity for $d\rho/dT < 0$ to $d\rho/dT > 0$ between y = 0.4 and y = 0.6 as shown in figure 4 (A).

## 4. CONCLUSIONS

We revealed that the simultaneous substitution of indium and selenium ($x = 1.0$ and $y \geq 0.6$) induces metallic behavior and a decrease in resistivity for $Co_3Sn_{2-x}In_xS_{2-y}Se_y$ without suppression of the Seebeck coefficient. From the electron band calculations, assuming the strong site preference for indium substitution ($Sn1_{In92}$), the selenium doping does not contribute any orbital levels of selenium at the Fermi level. However, when the selenium content is over $y = 0.5$, the total crystallographic energy shows that a higher indium occupancy at the Sn2 site is more stable. Therefore, the selenium doping suppresses the site preference, resulting in an increase in the indium occupancy at Sn2 site. This is one of the possible explanations for the observed metallic conductivity in Se-doped $Co_3SnInS_2$.

ACKNOWLEDGMENTSF

This work was partially supported by the research grants from Keio University, the Keio Leadingedge Laboratory of Science and Technology (KLL)

Figure caption

Figure 1. (a) Crystal structure of $Co_3Sn_2S_2$. (b) Kagome lattice composed of Co. (c) Lattice site location of Sn1 and Sn2.

Figure 2. XRD pattern for $Co_3SnInS_{1-y}Se_y$.

Figure 3. Lattice parameters of the a- and c-axes as a function of selenium concentration (y). Inset shows the detected composition of selenium and indium as a function of y obtained from the EDX results.

Figure 4. Resistivity (ρ), Seebeck coefficient (α), thermal conductivity (κ), figure of merit (ZT) versus temperature (T) for bulk $Co_3InSnS_{2-y}Se_y$ (0<y<1).

Figure 5. Densities of states for $Co_3SnInS_2$. The shadowed area shows the partial DOS of Co. (A): Indium occupies all of Sn1 sites ($Sn1_{In100}$). (B): Indium occupies 92 % of Sn1 sites and 8 % of Sn2 sites ($Sn1_{In92}$). (C): Indium occupies 75 % of Sn1 sites and 25 % of Sn2 sites ($Sn1_{In75}$).

Figure 6. Densities of states for $Co_3InSnS_{2-y}Se_y$. The shaded area shows the partial DOS of Co. (A), (B), (C): Indium occupies 92 % of Sn1 sites and 8 % of Sn2 sites ($Sn1_{In92}$) and y = 0.125, 0.5 and 1.0 respectively. (D): Indium occupies 75 % of Sn1 sites and 25 % of Sn2 sites ($Sn1_{In75}$) and y = 0.5.

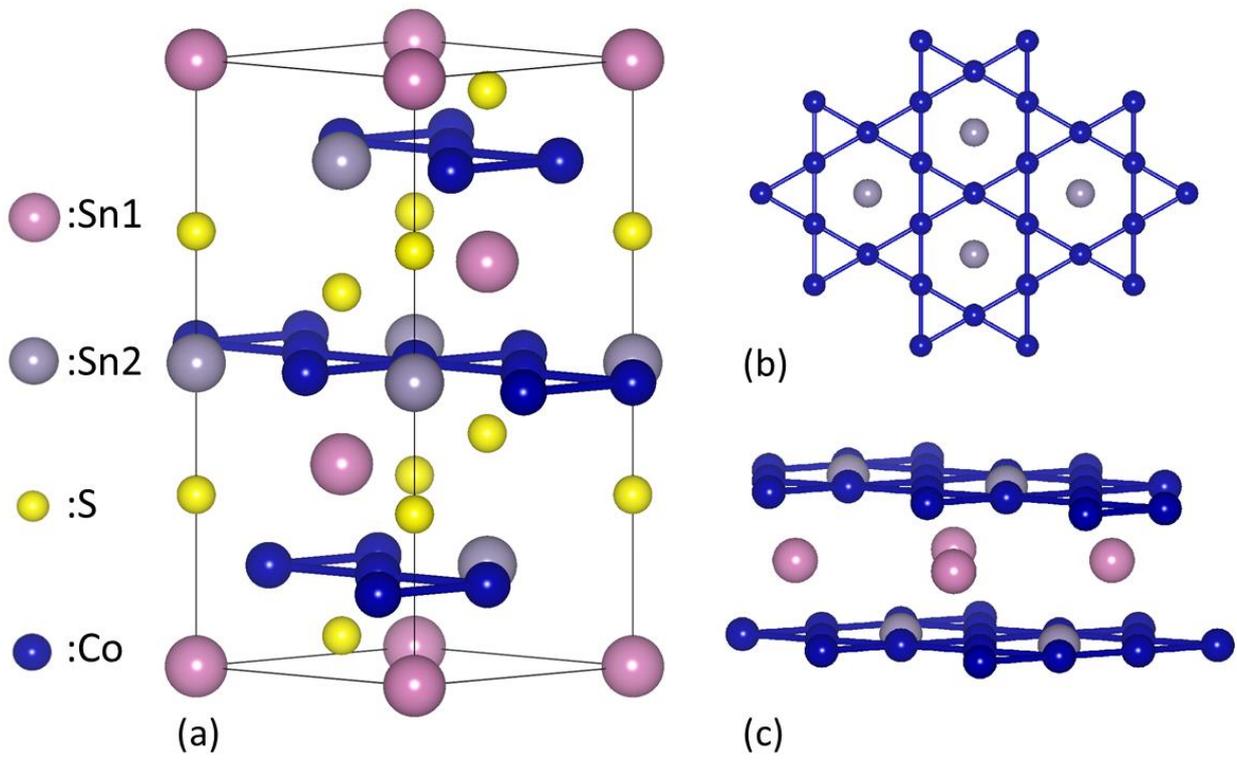

Fig. 1

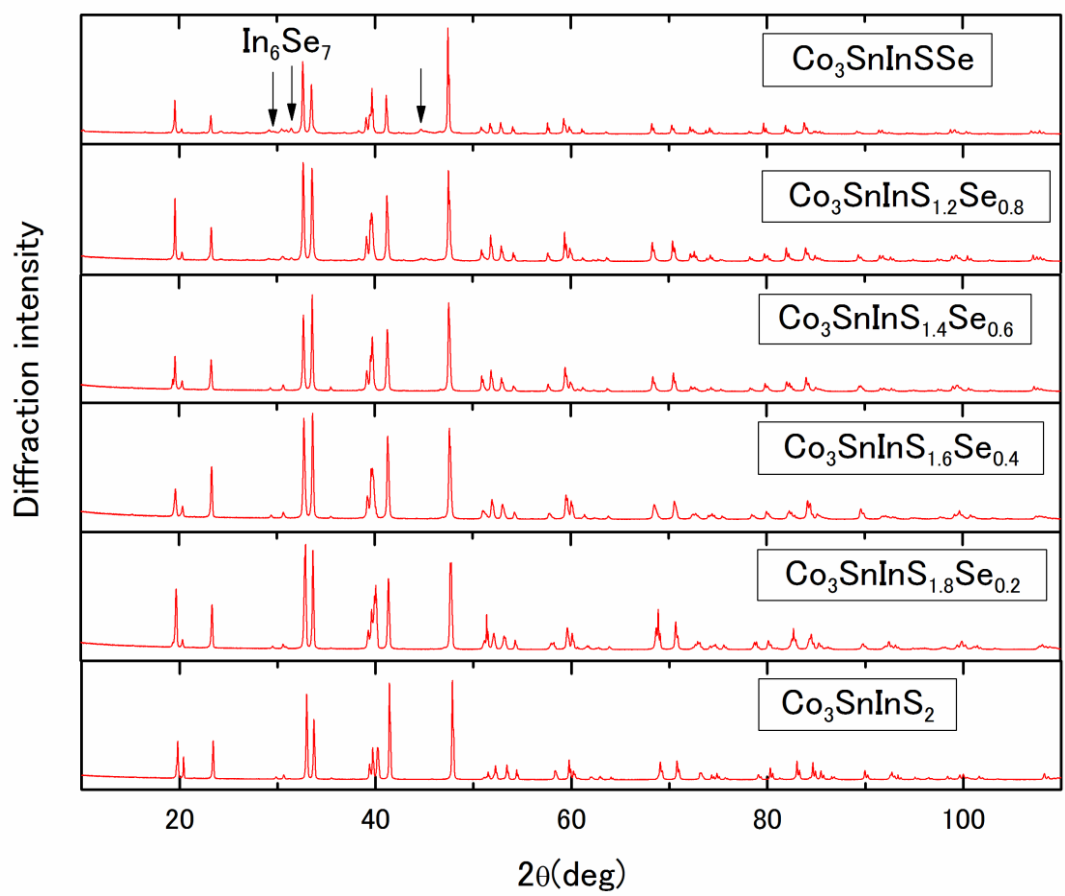

Fig. 2

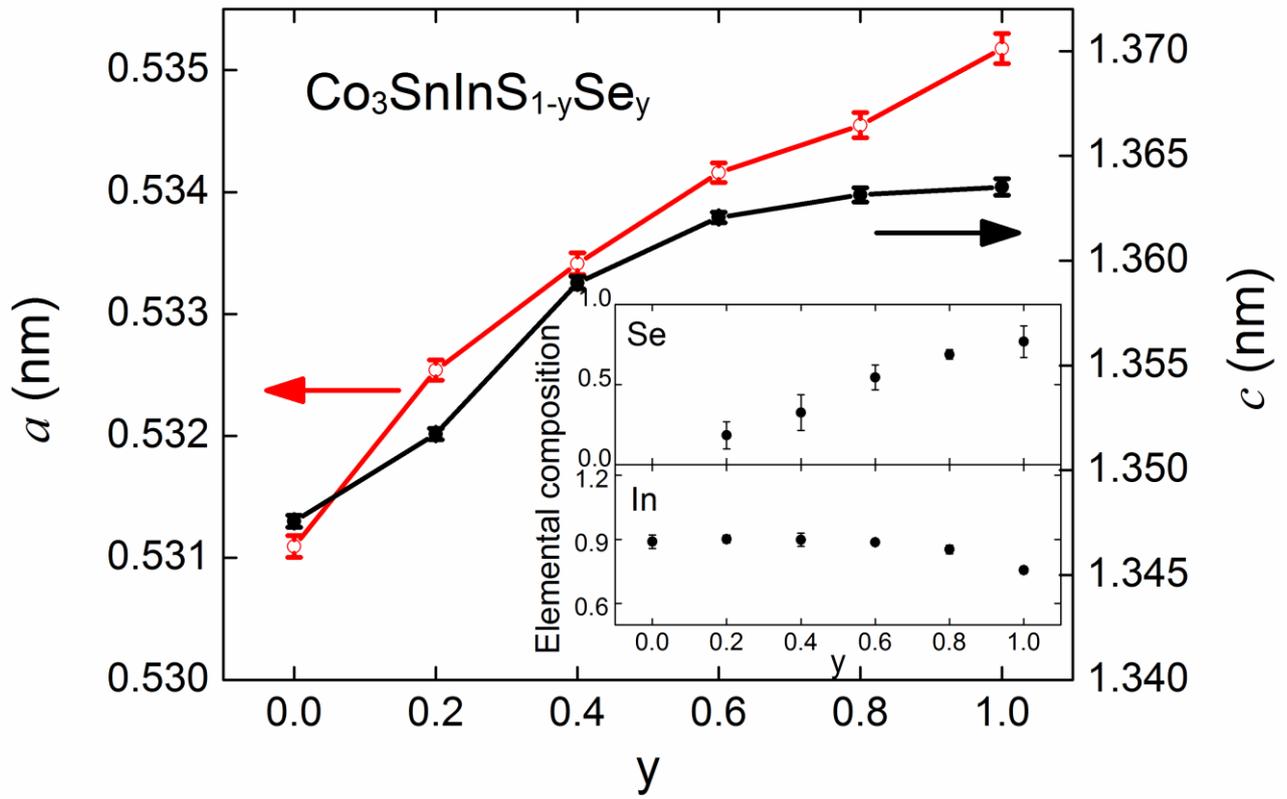

Fig. 3

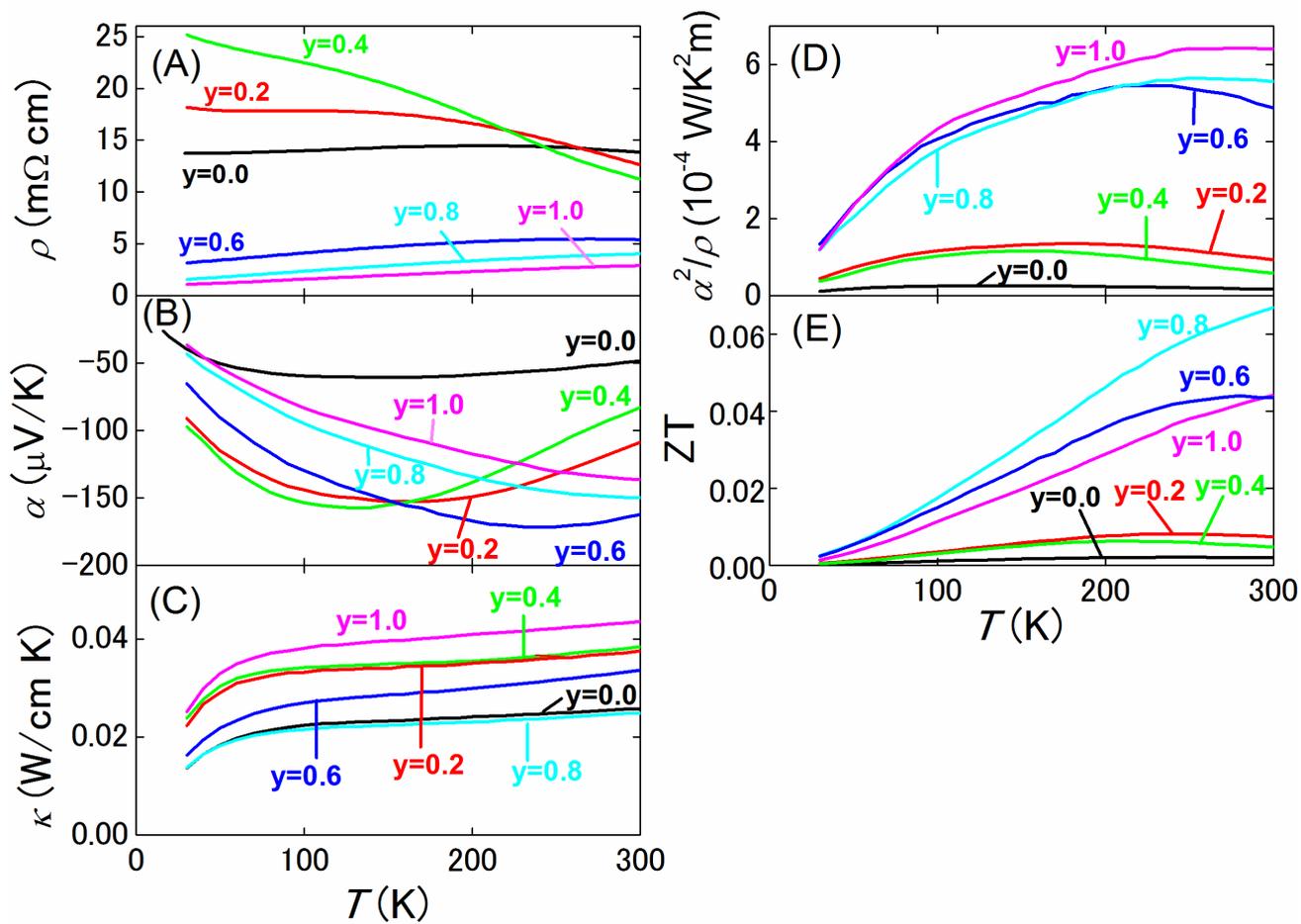

Fig. 4

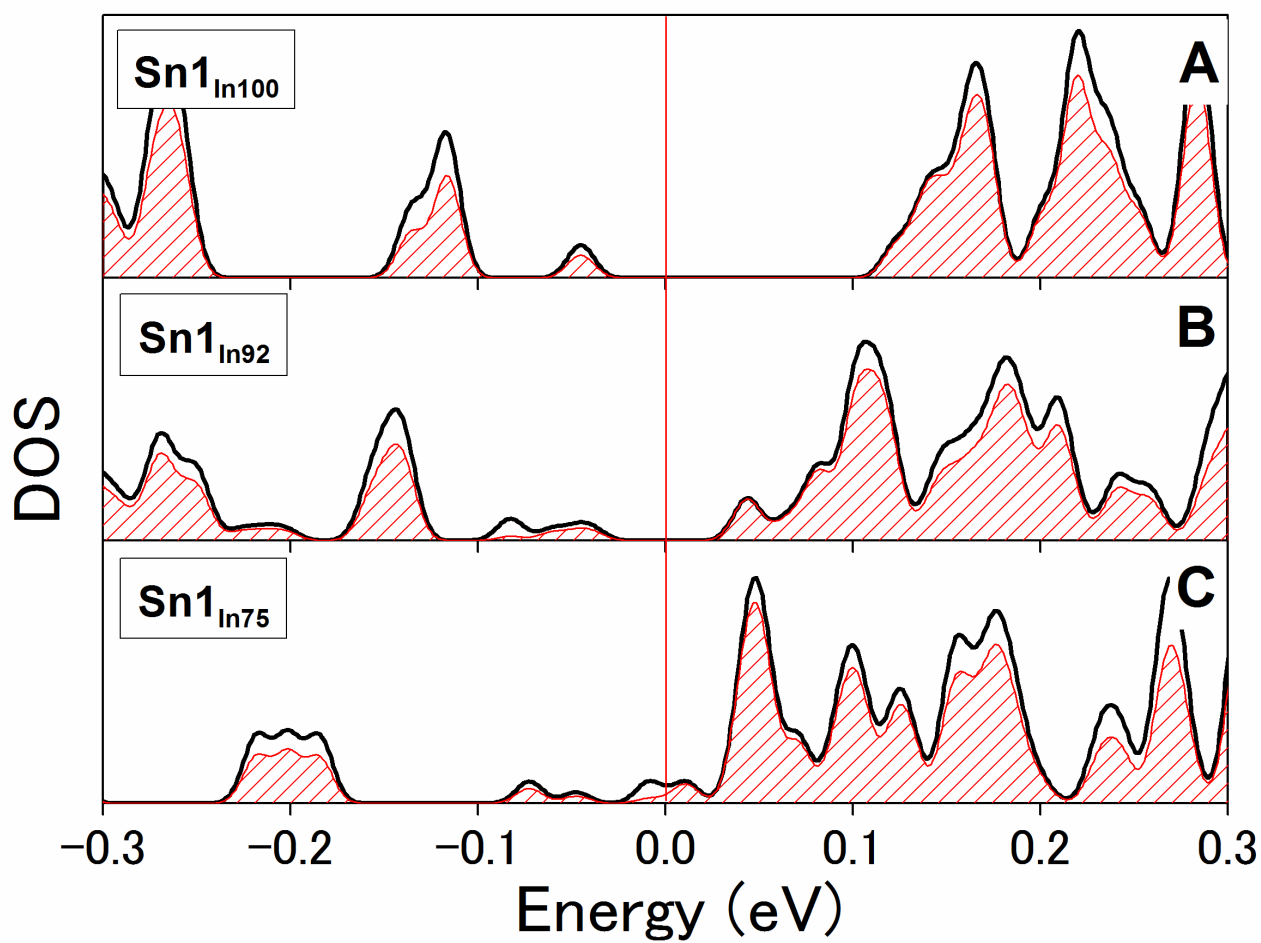

Fig. 5

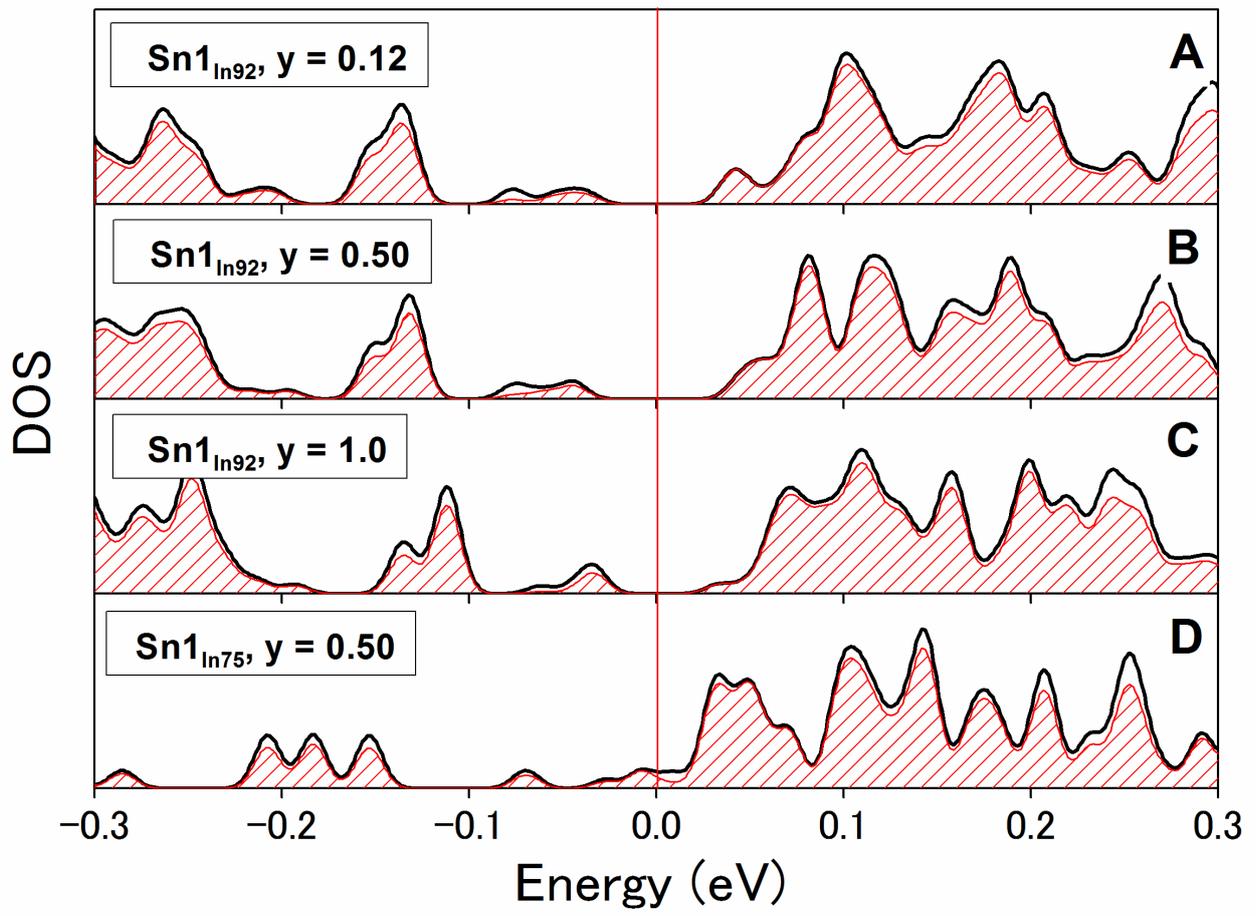

Fig. 6